\documentclass[aps,prb,reprint,showkeys,superscriptaddress,floatfix,twocolumn]{revtex4-2}
\usepackage[colorlinks,citecolor=blue,urlcolor=blue,bookmarks=false,hypertexnames=true]{hyperref} 
\usepackage[b]{esvect}
\usepackage{amsfonts}
\usepackage{amsmath}
\usepackage{amssymb}
\usepackage{subcaption}
\usepackage{graphicx}
\usepackage{caption}
\usepackage{braket}
\usepackage{comment}

\begin{document}
\title{Breakdown of superdiffusion in perturbed quantum integrable spin chains and ladders}
\author{Kevin Wang}
\affiliation{Department of Physics, University of California, Berkeley, California 94720, USA}
\author{Joel E. Moore}
\affiliation{Department of Physics, University of California, Berkeley, California 94720, USA}
\affiliation{Materials Sciences Division, Lawrence Berkeley National Laboratory, Berkeley, California 94720, USA}
\date{\today}
\begin{abstract}
Superdiffusive transport with dynamical exponent $z=3/2$ has been firmly established at finite temperature for a class of integrable systems with a non-abelian global symmetry $G$. On the inclusion of integrability-breaking perturbations, diffusive transport with $z=2$ is generically expected to hold in the limit of late time. Recent studies of the classical Haldane-Ishimori-Skylanin model have found that perturbations that preserve the global symmetry lead to a much slower timescale for the onset of diffusion, albeit with uncertainty over the exact scaling exponent.
That is, for perturbations of strength $\lambda$, the characteristic timescale for diffusion goes as $t_*\sim \lambda^{-\alpha}$ for some $\alpha$. Using large-scale matrix product state simulations, we investigate this behavior for perturbations to the canonical \textit{quantum} model showing superdiffusion: the $S=1/2$ quantum Heisenberg chain. We consider a ladder configuration and look at various perturbations that either break or preserve the $SU(2)$ symmetry, leading to scaling exponents consistent with those observed in one classical study \cite{mccarthy_slow_2024}: $\alpha=2$ for symmetry-breaking terms and $\alpha=6$ for symmetry-preserving terms. We also consider perturbations from another integrable point of the ladder model with $G=SU(4)$ and find consistent results. Finally, we consider a generalization to an $SU(3)$ ladder and find that the $\alpha=6$ scaling appears to be universal across superdiffusive systems when the perturbations preserve the non-abelian symmetry $G$.
\end{abstract}

\maketitle

\section{Introduction}
The quantum Heisenberg model (QHM) is a paradigm of quantum magnetism and has been studied extensively since its introduction in 1928. It is a prototypical model for the study of quantum corrections to classical magnetic systems and is a good zeroth-order description for many materials that can be modeled by localized magnetic moments. In one dimension, its physics is extremely rich. For example, for the antiferromagnetic QHM, it is known that the half-integer spin chains ($S=1/2,3/2,\ldots$) are gapless, with the low-temperature physics described by a Luttinger liquid, while the integer spin chains ($S=1,2,\ldots$) are gapped~\cite{haldane_gap}. The $S=1$ Haldane phase is a prototypical example of a symmetry protected topological  phase.

The ground state and thermodynamics of the $S=1/2$ spin chain is famously exactly solvable through the techniques of the Bethe ansatz; it falls under a class of quantum models known to be integrable, in which there are an infinite number of locally conserved quantities. This behavior is very different from typical systems, which are described by chaotic dynamics and have finitely many conserved quantities such as energy, momentum, and spin. Given that so many properties of the one-dimensional (1D) QHM are exactly solvable, it is perhaps surprising that only in recent years did research uncover novel behavior for spin transport at high temperatures. The purpose of the present work is to understand how that novel behavior is expected to appear in realistic systems including non-negligible perturbations to the ideal QHM.

While the low-energy physics of transport in 1D quantum systems largely focuses on the gapped vs.~gapless nature of excitations, high-temperature transport at long times and length scales can be described by an emergent hydrodynamic theory. In this framework, one assumes that the system has locally equilibrated to a Gibbs ensemble specified by chemical potentials for each conserved quantity, in which case one studies transport through a set of coupled partial differential equations written solely in terms of these locally conserved densities. Unless a current is explicitly conserved (such as in Galilean systems), the transport behavior is typically expected to be diffusive, with dynamical exponent $z=2$ characterizing the scaling of time with space $t\sim x^z$. Spin transport in the $S=1/2$ Heisenberg chain, however, displays an anomalous exponent of $z=3/2$ that is \textit{superdiffusive} \cite{znidaric_spin_2011,ljubotina_class_2017,ljubotina_kardar-parisi-zhang_2019}. This value for the critical exponent is known to characterize the universality class for a stochastic partial differential equation introduced by Kardar, Parisi, and Zhang (KPZ) to describe the growth of interfaces in contexts as varied as flame fronts, tumors, etc. \cite{kardar_dynamic_1986}.

\begin{figure*}[t]
    \centering
    \includegraphics[width=0.8\textwidth]{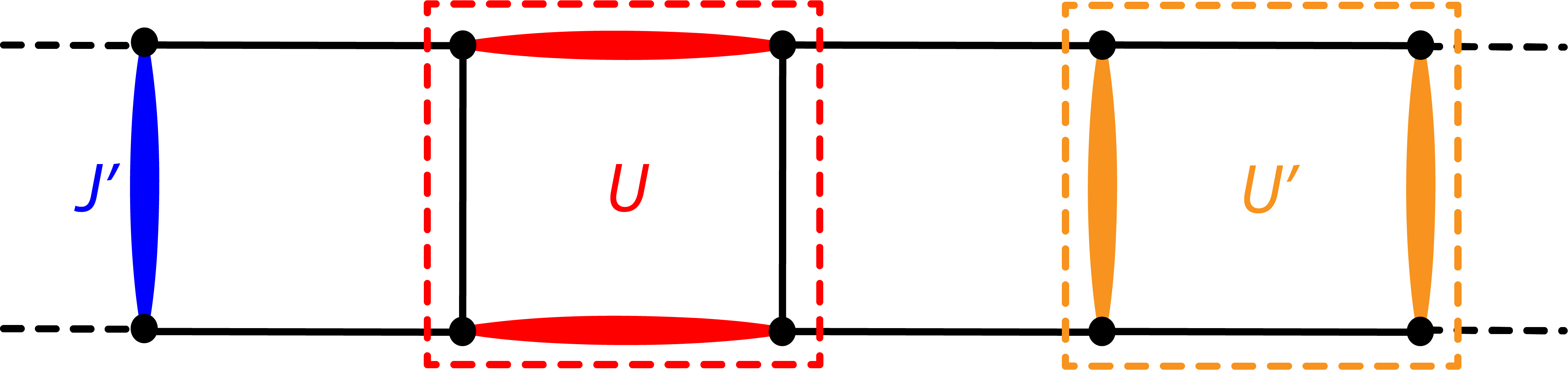}
    \caption{\label{fig:hamiltonian}Interactions added to the unperturbed set of two Heisenberg chains as considered in Eq. \ref{hamiltonian}. They consist of a Heisenberg interaction across the rungs of the ladder (blue), as well as two biquadratic terms which are products of two Heisenberg interactions (red, orange). The $J'$ and $U'$ terms explicitly break the $SU(2)$ symmetry of the individual chains, while the $U$ term does not. At the $SU(4)$ integrable point, both biquadratic perturbations break the symmetry, while the rung coupling reduces to a chemical potential.}
\end{figure*}

Developing a hydrodynamic description of quantum integrable systems is complicated by the existence of an infinite set of conserved quantities. Recently, the theory of generalized hydrodynamics (GHD) has been developed to address this limitation \cite{bertini_transport_2016,castro-alvaredo_emergent_2016,doyon_lecture_2020,doyon_generalized_2023}. Under this framework, one can transform the set of coupled equations from the basis of conserved quantities to a basis of quasiparticle densities, for which thermodynamic properties are known from the technology of the Bethe ansatz. The result for the easy-plane spin chain is a kinetic equation of the same form as appears in some classical integrable models, such as for a gas of solitons in the nonlinear Schr\"odinger equation.~\cite{elkamchatnov} Transport quantities can then be calculated in terms of these quasiparticles, each propagating ballistically and carrying some amount of charge dressed by its collisions with other quasiparticles.

The Heisenberg point is considerably more challenging, but a theory for spin superdiffusion has been developed in which the quasiparticles contributing to spin transport are large solitonic wavepackets of Goldstone modes \cite{gopalakrishnan_kinetic_2019,gopalakrishnan_anomalous_2019,bulchandani_kardar-parisi-zhang_2020,de_nardis_superdiffusion_2020}. The physics underlying this anomalous exponent is now believed to be understood, and similar behavior is expected to apply to other integrable systems with short-range interactions and non-abelian symmetry \cite{dupont_universal_2020,ye_universal_2022,ilievski_superuniversality_2021}. (Here we follow previous authors in using ``non-abelian'' to refer to cases where the related Lie group symmetry is non-abelian, not cases obtained by combining $U(1)$ with discrete symmetries in a non-abelian way.)

In physical systems, this behavior has been observed in neutron scattering experiments of KCuF$_3$, a material well described by the 1D $S=1/2$ antiferromagnetic QHM \cite{scheie_detection_2021}, as well as in cold-atom quantum simulators \cite{wei_quantum_2022} and quantum computers \cite{google_quantum}. However, experimental systems are never perfectly integrable as integrability-breaking perturbations are inevitably present. In this case, the transport behavior of conserved densities at long times is expected to become diffusive, with exponent $z=2$. A central question is then how to characterize the effects of introducing integrability-breaking perturbations to the model.

In this paper, we focus our investigation on a ladder configuration as a correction to the 1D integrable model, partially motivated by quasi-1D systems like KCuF$_3$. Recent works \cite{mccarthy_slow_2024,mcroberts_parametrically_2024} have explored the effect of integrability-breaking terms in \textit{classical} integrable models, namely the Haldane-Ishimori-Sklyanin (HIS) chain \cite{haldane_integrable_1982,ishimori_integrable_1982,sklyanin_1982}. A central result of these works is the observation of a much slower onset of diffusion when considering perturbations that preserve the non-abelian symmetry of the integrable model. We explore this behavior in a \textit{quantum} setting using state-of-the-art matrix product state (MPS) simulations. First, we consider adding a variety of perturbations which either preserve or break the $SU(2)\times SU(2)$ symmetry of the unperturbed system of two decoupled Heisenberg chains. We then consider a different integrable point of this ladder model \cite{wang_exact_1999} which has emergent $SU(4)$ symmetry, and verify that the same set of perturbations displays different behavior due to the modified non-abelian symmetry underlying the integrability. Finally, motivated by evidence~\cite{dupont_universal_2020} and theoretical arguments~\cite{ilievski_superuniversality_2021} made for the universality of superdiffusion across a broad class of non-abelian integrable models, we generalize the system to a ladder with $SU(3)\times SU(3)$ symmetry and confirm that there is a similar universality underlying the slow onset of diffusion for symmetric perturbations.

\section{Spin transport}
There are various approaches to extracting the anomalous dynamical exponent $z$ for a given spin model. The method we follow is that of relaxation from an inhomogeneous quench \cite{ljubotina_spin_2017}. We initialize a mixed state as
\begin{equation}
\label{initial state}
    \rho_\mu(t=0)\sim \exp(\mu S^z_\text{left})\otimes\exp(-\mu S^z_\text{right})
\end{equation}
where $S^z_\text{left}$ ($S^z_\text{right}$) is the total $S^z$ charge in the left (right) half of the chain. A nonzero $\mu$ describes an infinite temperature state at finite, inhomogeneous chemical potential (more precisely, the ratio of a spin chemical potential to temperature is held finite), and the limit $\mu\rightarrow0$ recovers the infinite temperature mixed state. A natural measure of transport for this initial condition is the magnetization transfer between the two halves of the system, which can be measured as
\begin{equation}
\label{mag transfer}
    \Delta s(t) \equiv \langle S^z_\text{left}(t)\rangle_\mu-\langle S^z_\text{left}(0)\rangle_\mu
\end{equation}
The scaling of the magnetization transfer, after an initial local equilibration time, is expected to follow $\Delta s(t) \sim t^{1/z}$. This is indeed found to hold with $z=3/2$ for the 1D quantum Heisenberg model \cite{ljubotina_spin_2017}. However, with integrability-breaking terms added to the Hamiltonian, diffusive behavior ($z=2$) should prevail in the asymptotic limit. As a result, one expects that for small perturbations from an integrable model, there is an interpolation from an initial, KPZ-like exponent $z=3/2$ to the asymptotic diffusive behavior $z=2$. Indeed, recent numerical investigations \cite{mccarthy_slow_2024, mcroberts_parametrically_2024} have found this to be true in classical models, namely for the perturbed integrable HIS chain.

\begin{figure*}[t]
    \centering
    \includegraphics[width=\textwidth]{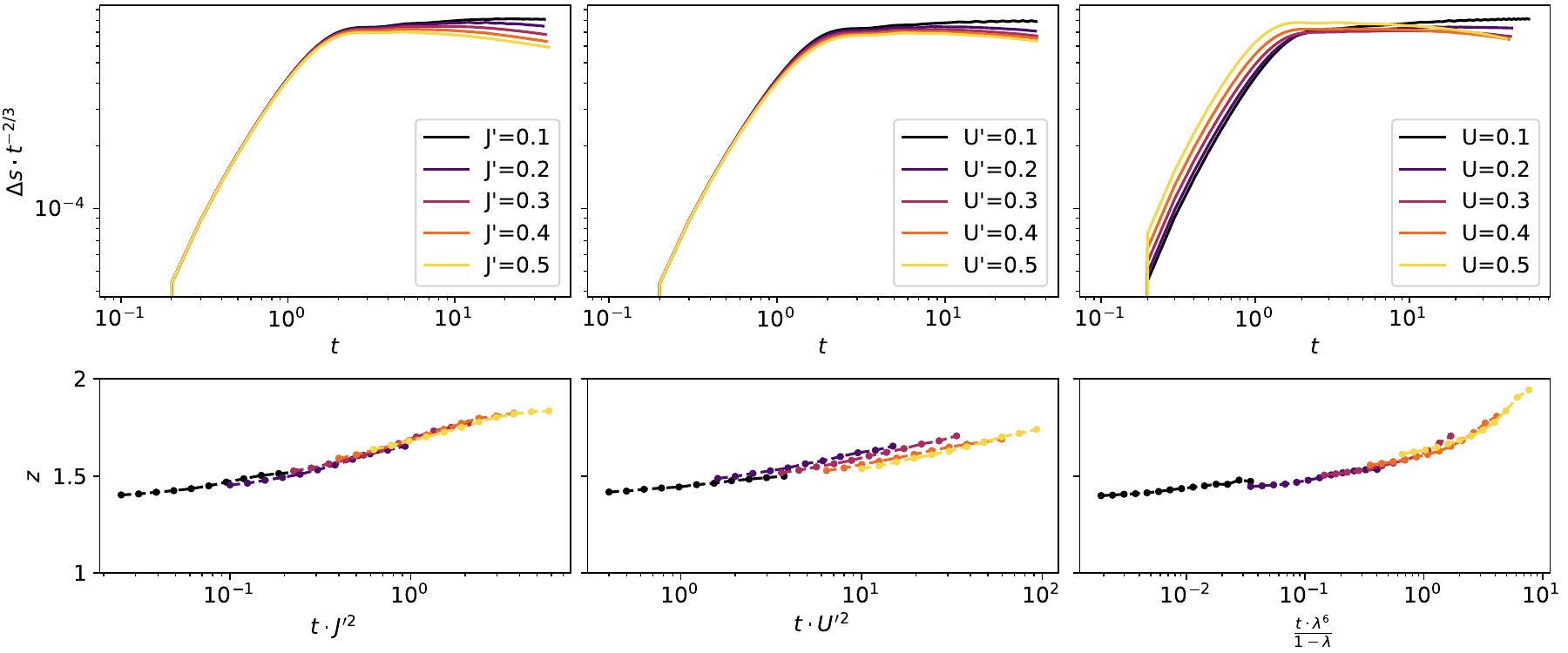}
    \caption{\label{fig:su2}Onset of diffusion due to perturbations from decoupled Heisenberg chains. Top row: magnetization transfer $\Delta s(t)$ as defined in Eq. \ref{mag transfer} with $\mu=0.001$, rescaled by $t^{-1/z}$ to highlight deviations from superdiffusive behavior over time. Bottom row: dynamical exponent as a function of time $z(t)$ extracted by fitting to a power law over local windows in time. The time $t$ is scaled by different powers of the perturbation strength according to whether or not the non-abelian $SU(2)\times SU(2)$ symmetry of the two integrable chains is preserved. The $J'$ (left) and $U'$ (middle) perturbations do not preserve this symmetry, and the scaling collapse follows that predicted by Fermi's golden rule. The $U$ (right) perturbation does preserve this symmetry, and we find a slower scaling $t_*\sim\lambda^{-6}$ where $\lambda\equiv\frac{U}{1+U}$ in accordance with the parametrization considered in \cite{mccarthy_slow_2024}.}
\end{figure*}

We verify that these results also hold for \textit{quantum} systems. In particular, we take our unperturbed integrable model to be two decoupled spin-1/2 Heisenberg chains arranged in a ladder configuration. Of course, with no coupling between the chains, transport properties should match that of an individual chain. However, there are a variety of interesting terms that can be added to the Hamiltonian that lead to different behaviors for the scaling of the onset of diffusion with the perturbation strength. Concretely, we consider the model
\begin{equation} \label{hamiltonian}
\begin{split}
    H &=J\sum_{\langle i,j \rangle}\left(\vec{S}^{(1)}_{i}\cdot\vec{S}^{(1)}_{j}+\vec{S}^{(2)}_{i}\cdot\vec{S}^{(2)}_{j}\right)\\
    &+J'\sum_i\left(\vec{S}^{(1)}_{i}\cdot\vec{S}^{(2)}_{i}\right)\\
    &+4U\sum_{\langle i,j \rangle}\left(\vec{S}^{(1)}_{i}\cdot\vec{S}^{(1)}_{j}\right)\left(\vec{S}^{(2)}_{i}\cdot\vec{S}^{(2)}_{j}\right)\\
    &+4U'\sum_{\langle i,j \rangle}\left(\vec{S}^{(1)}_{i}\cdot\vec{S}^{(2)}_{i}\right)\left(\vec{S}^{(1)}_{j}\cdot\vec{S}^{(2)}_{j}\right)
\end{split}
\end{equation}
The superscript 1 (2) indicates an operator acting on the top (bottom) chain of the ladder (Note: the $S^z_\text{left}$ and $S^z_\text{right}$ operators in Eqs. \ref{initial state} and \ref{mag transfer} are generalized accordingly to the charge in the two halves of the \textit{ladder}). The $J'$ terms introduce Heisenberg interactions along the rungs of the ladder, and the $U$ and $U'$ terms are (total) $SU(2)$-invariant biquadratic interactions on plaquettes (Fig. \ref{fig:hamiltonian}). For simplicity, we set $J=1$ from here on out and focus on the scaling of diffusion onset with the various perturbation strengths $J',U,U'$. Beyond the regime of decoupled chains ($J'=U=U'=0$), there are two other integrable points for particular values of $U$ and $U'$, for which exact solutions have been determined through Bethe ansatz techniques \cite{wang_exact_1999}. In addition to the regime in the vicinity of two decoupled chains, we will also look at perturbations from one of the other integrable points ($U=J,U'=0$), where the ladder can be coarsegrained into a \textit{chain} with emergent $SU(4)$ symmetry. Finally, to probe the universality of diffusion timescales from symmetric perturbations, we analyze a generalization of the $SU(2)$ ladder considered above to an $SU(3)$ ladder.

\begin{figure*}[t]
    \centering
    \includegraphics[width=\textwidth]{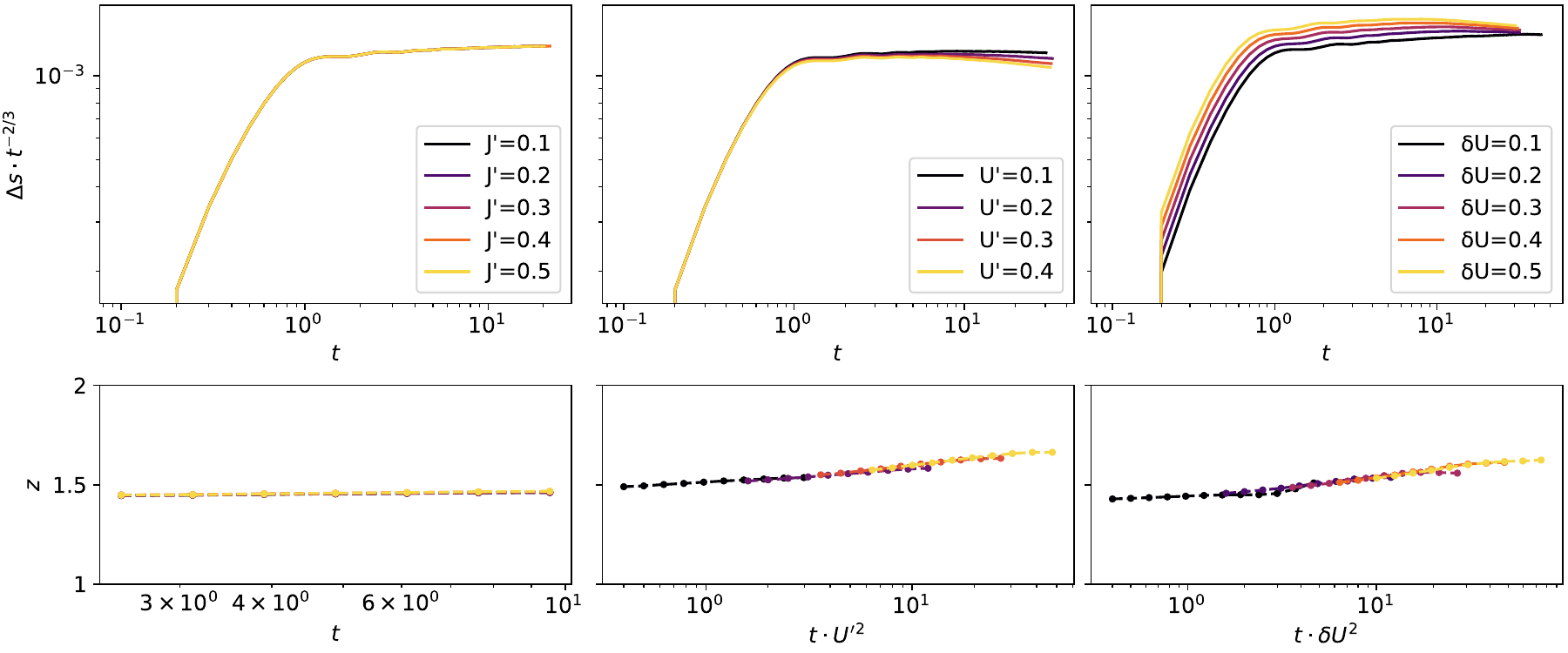}
    \caption{\label{fig:su4}Onset of diffusion due to perturbations from the $SU(4)$ integrable point $(U=J)$. The $J'$ (left) perturbation now reduces to a chemical potential for one of the Cartan charges of $SU(4)$, preserving integrability. Therefore, superdiffusion persists. Both $U'$ (middle) and $\delta U\equiv U-J$ (right) break $SU(4)$, and we find a scaling collapse in line with golden rule estimates.}
\end{figure*}

We simulate high-temperature spin transport using matrix product states \cite{schollwock_density-matrix_2011} with the purification method via ancilla sites \cite{verstraete_matrix_2004}. We construct the initial state (\ref{initial state}) by evolving an infinite temperature MPS in imaginary time by a domain wall Hamiltonian, to a chemical potential of $\mu=0.001$. All simulations are conducted for ladders of length $L=64$ and a maximum bond dimension of $\chi=512$. For the spin-1/2 system considered in the first section, the state is time-evolved using the time dependent variational principle (TDVP) \cite{haegeman_time-dependent_2011,haegeman_unifying_2016} with a time step of $\delta t=0.1$. The ancilla sites are time-evolved in reversed time to reduce the growth of entanglement \cite{karrasch_finite-temperature_2012}. The $SU(3)$ ladder considered later is less efficiently simulable via TDVP due to the large bond dimension of the matrix product operator (MPO) representing the Hamiltonian. For that system, we time-evolve using the standard time-evolving block decimation (TEBD) approach \cite{vidal_efficient_2004} with parallelization \cite{urbanek_parallel_2016}. An optimal fourth-order Trotter-Suzuki decomposition is chosen \cite{barthel_optimized_2020}, also with time step $\delta t=0.1$, and quantum number conservation is enforced for both of the $SU(3)$ Cartan charges.

\subsection{Perturbations from decoupled Heisenberg chains $(J'=U=U'=0)$}
Whether or not the perturbation preserves the symmetry of the Hamiltonian is an important consideration in determining the behavior of the non-integrable model \cite{de_nardis_stability_2021, mccarthy_slow_2024}. In particular, $SU(2)$-symmetric perturbations lead to a much slower onset of diffusion, scaling as $t_*\sim\lambda^{-6}$, where $\lambda$ characterizes the strength of the perturbation. When the symmetric perturbations are \textit{noisy}, which breaks total energy conservation, they still preserve a (logarithmically) divergent diffusion constant $D(t)\sim \log t$, while for symmetric, \textit{static} perturbations, preserving both energy and quasimomentum, the scattering matrix elements are more strongly restricted by selection rules, leading to continued KPZ-like behavior in the diffusion constant $D(t)\sim t^{1/3}$ within accessible timescales. On the other hand, perturbations that do not preserve the symmetries of the Hamiltonian generically follow predictions from golden rule estimates where the scattering rate simply goes as the perturbation strength squared $\Rightarrow t_*\sim\lambda^{-2}$.

The derivation of these scaling exponents builds off of the kinetic theory of string quasiparticles that provides a simple picture for superdiffusion \cite{de_nardis_stability_2021, mccarthy_slow_2024}. We reproduce those arguments here for clarity. The modern theory of generalized hydrodynamics allows one to express the transport behavior of integrable systems in terms of ballistically moving quasiparticles carrying a charge dressed by collisions with other quasiparticles. Under this framework, spin transport in the isotropic Heisenberg chain at half-filling is represented as a sum over contributions from all quasiparticles. At half-filling, all strings carry zero net charge at long times since collisions with larger strings will neutralize the charge current, leading to zero Drude weight. However, one can still characterize the timescale at which the collisions occur, i.e. how long current is initially carried for, using the thermodynamic Bethe ansatz~\cite{takahashi_1999}. At infinite temperature, the density of string excitations of length $s$ is $\rho_s\sim s^{-3}$. Therefore, for any string of length $s$, the density of larger strings goes as $s^{-2}$. If we multiply this by the dressed velocity of the string $v_s^{\text{dr}}\sim s^{-1}$, we get an intrinsic screening rate of $\Gamma_s^0\sim s^{-3}$. One can then derive the superdiffusive exponent from the Kubo formula for current-current correlations.

If we now wish to characterize integrability-breaking perturbations, a new scattering rate $\Gamma_{s,\lambda}$ is introduced in the sum, which to lowest order can be calculated from Fermi's golden rule. If this scattering rate has no $s$-dependence $\Gamma_{s,\lambda}\sim \lambda^2$, then we should simply expect $t_*\sim\lambda^{-2}$. However, perturbations that preserve the $SU(2)$ symmetry of the Hamiltonian will preserve the total magnetization $s$. Since these large strings can be identified as solitonic wavepackets of Goldstone modes, with energy scaling as $\varepsilon_s\sim s^{-1}$, any matrix element of a symmetry-preserving perturbation is expected to be bounded by this scaling, i.e. $\Gamma_{s,\lambda}\lesssim \lambda^2s^{-2}$ \cite{de_nardis_stability_2021}. This scattering rate dominates the intrinsic screening when $\lambda^2s^{-2} \gtrsim s^{-3}\Rightarrow\lambda^6\gtrsim s^{-3}$. On the other hand, the time at which the screening occurs is given by $s^{-3}\gtrsim 1/t$, which leads to the scaling $t_*\sim\lambda^{-6}$ \cite{mccarthy_slow_2024}.

In considering perturbations from the set of two decoupled Heisenberg chains, one might naively expect all of the extra terms in Eq. \ref{hamiltonian} to be regarded as ``symmetric perturbations,'' as they all preserve the total $SU(2)$ symmetry of the model. However, it is really the $SU(2)\times SU(2)$ symmetry, i.e. the independent conservation of magnetization for each chain, that is important in considering the allowed scattering mechanisms. Perturbations that do not preserve this symmetry allow for transfers of magnetization from one chain to the other. Therefore, the magnetization $s$ of a string quasiparticle in a particular chain is not protected, and the Goldstone physics that would lead to the $s$-dependent scattering rate bound $\Gamma_s\lesssim s^{-2}$ does not hold in this case \cite{de_nardis_stability_2021, mccarthy_slow_2024}. This is true for the $J'$ and $U'$ perturbations in Eq. \ref{hamiltonian}. As shown in Fig. \ref{fig:su2}, we find an onset of diffusion that matches generic golden rule expectations for these perturbations. The $J'$ perturbation is of physical relevance as an approximation of physical systems which are not perfect one-dimensional chains.

On the other hand, when we consider the biquadratic $U$ perturbations, the $SU(2)\times SU(2)$ symmetry is protected, and we indeed see a much slower onset of diffusion, consistent with the static perturbations considered in Ref. \cite{mccarthy_slow_2024}. In that paper, the authors chose to parameterise the integrability-breaking with the Hamiltonian $H=(1-\lambda)H_0+\lambda H'$, and found the time for diffusion to occur to scale as $t_*\sim\lambda^{-6}$. We also find this to be true for $\lambda=\frac{U}{1+U}$ after rescaling our Hamiltonian by $(1-\lambda)$ to match their form. In the limit $\lambda\rightarrow0$, the difference between the two scaling parameters is negligible. Fig. \ref{fig:su2} shows a clear scaling collapse of $\frac{t\cdot \lambda^6}{1-\lambda}$. We note that this perturbation is in the same regime as the symmetric, static perturbations considered previously \cite{de_nardis_stability_2021}.

\subsection{Perturbations from $SU(4)$ point $(U=J, J'=U'=0)$}
Moving away from the integrable point of decoupled Heisenberg chains, we now look at the behavior of perturbations from the $SU(4)$ point of the spin ladder. At this particular point, the model simplifies into a sum of permutation operators between nearest neighbors, where the two sites on each rung are combined into a single 4-dimensional Hilbert space, effectively forming an $SU(4)$ \textit{chain}. Previous numerical studies have observed anomalous transport for this model \cite{znidaric_magnetization_2013}, although the dynamical exponent was estimated to be $z=5/4$ from a general conjecture of $z=(N+1)/N$ for $SU(N)$ permutation models (the Heisenberg chain being an $N=2$ permutation model). However, arguments have since been made for a ``superuniversality'' of superdiffusive $z=3/2$ transport of Noether charges in integrable models protected by a global symmetry defined by a non-abelian, simple Lie group $G$ \cite{ilievski_superuniversality_2021}, supported by several numerical studies \cite{dupont_universal_2020, fava_spin_2020, prosen_macroscopic_2013, das_kardar-parisi-zhang_2019, krajnik_kardarparisizhang_2020, krajnik_integrable_2020}. We confirm that $z=3/2$ does hold for this model, and once again analyze how the onset of $z=2$ diffusion scales with perturbation strength.

Note first that the $J'$ rung interaction no longer breaks integrability as this term can now be interpreted as a chemical potential for one of the fifteen Noether charges of $SU(4)$, as noted in \cite{wang_exact_1999}. Adding this term reduces the symmetry of the Hamiltonian from $SU(4)$ to $U(1)\times SU(3)$ while preserving the integrability. In principle, one could choose two more commuting charges to form a Cartan subalgebra of $SU(4)$, and adding chemical potential terms for these charges would reduce the symmetry further into the so-called ``maximal torus'' $U(1)\times U(1) \times U(1) \times U(1)$ without breaking integrability. Superdiffusive transport for any of these charges $Q^{(n)}= \sum_i q^{(n)}_i$ still holds even in the presence of the chemical potentials, as is easily seen from the fact that the set of $Q^{(n)}$ by definition commute with both the Hamiltonian and each other, so that $q^{(n)}_i(t)=\exp{(iHt)}q^{(n)}_i\exp{(-iHt)}$ does not change, and therefore the expectation value in the infinite temperature Gibbs state is also unchanged (Fig. \ref{fig:su4}). Of course, this is no longer the case if we consider grand-canonical Gibbs states in which chemical potentials are included in the initial density matrix.

We continue to look at the transport of $S^z$ charge (the total $S^z$ charge on the two sites of a rung can be taken to be another Cartan charge of $SU(4)$). If we consider the two biquadratic perturbations, i.e. $\delta U\equiv U-J$ and $U'$, we see that neither of the perturbations now preserve the non-abelian $SU(4)$ symmetry. Therefore, we predict that both of these terms lead to the golden rule scaling of diffusion onset $t_*\sim\lambda^{-2}$ (Fig. \ref{fig:su4}). This naturally leads to the question as to whether we should still expect a slower $\lambda^{-6}$ scaling in other integrable non-abelian models besides the $SU(2)$ Heisenberg model, i.e. whether there is a corresponding ``superuniversality'' for this type of scaling. In models with larger symmetry groups $G$ of rank $r$, the generalized hydrodynamics approach to transport must account for the existence of $r$ different quasiparticle types with independent string lengths $s_a$, $a=1,\ldots,r$. If we introduce an integrability-breaking perturbation that preserves $G$, then each of the $s_a$ must be individually preserved, and the Goldstone physics of each quasiparticle type should in principle lead to the same $\lambda^{-6}$ scaling for the transport of any one of the charges.

\begin{figure}[t]
    \includegraphics[width=0.9\linewidth]{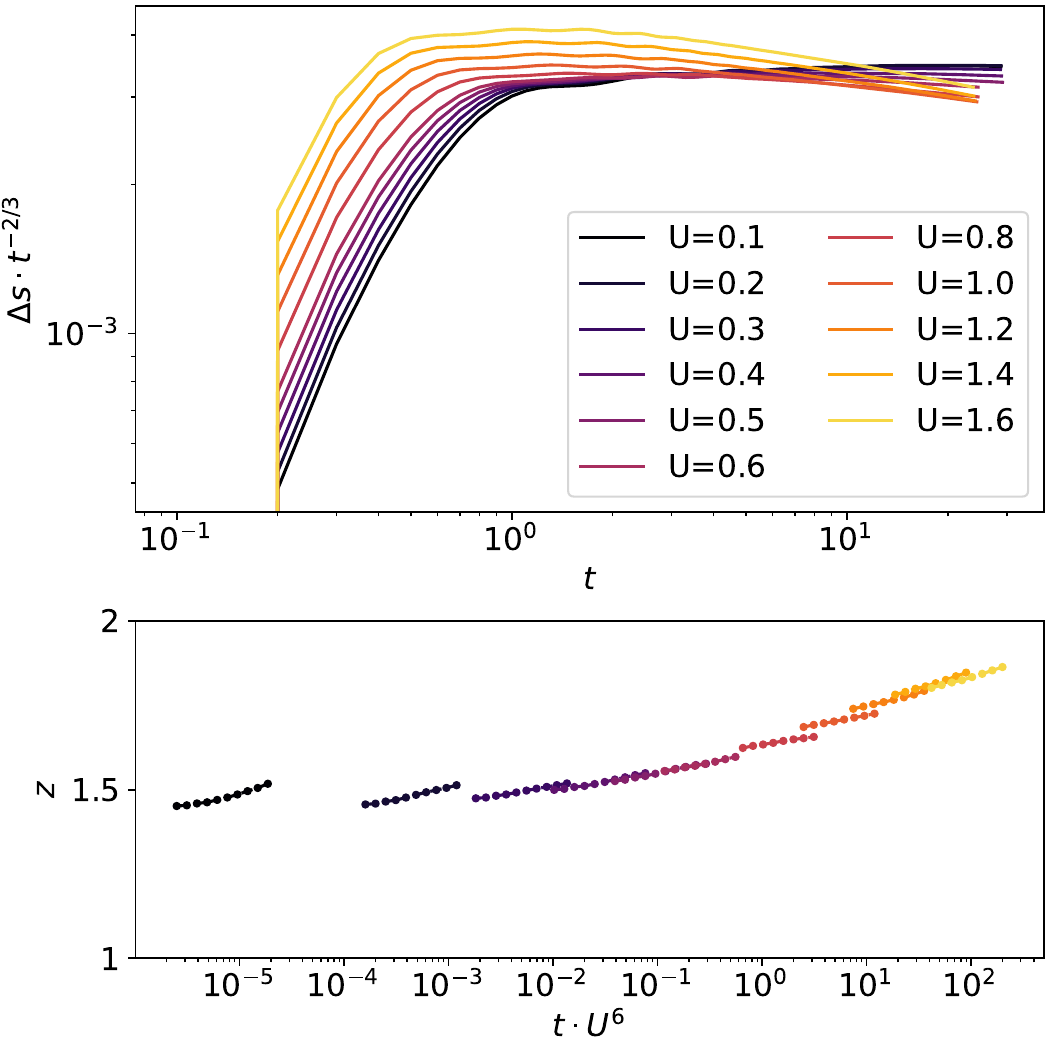}
    \caption{\label{fig:su3_U}Transport in a generalization of the model to an $SU(3)$ ladder, as specified in Eq. \ref{hamiltonian2}. The biquadratic perturbation preserves the non-abelian $SU(3)$ symmetry, again leading to a $t_*\sim U^{-6}$ scaling collapse, indicating the universality of slow diffusion onset if the non-abelian symmetry producing superdiffusion is protected.}
\end{figure}

\subsection{$SU(3)\times SU(3)$ Ladder}

To verify this behavior, we consider a minimal generalization of our system to an $SU(3)\times SU(3)$ ladder. That is, we consider a ladder geometry with a 3-dimensional local Hilbert space and a Hamiltonian given by

\begin{equation} \label{hamiltonian2}
\begin{split}
    H &=\sum_{\langle i,j \rangle}P^{(1)}_{i,j}+P^{(2)}_{i,j}+U\left(P^{(1)}_{i,j}\cdot P^{(2)}_{i,j}\right)
\end{split}
\end{equation}
where the operator $P^{(1)}_{i,j}$ ($P^{(2)}_{i,j}$) permutes the states between site $i$ and $j$, on the top (bottom) chain of the ladder. They are natural $SU(3)$-preserving generalizations of the Heisenberg couplings in the $SU(2)$ case. The $U$ term is therefore a generalization of the corresponding biquadratic term in Eq. \ref{hamiltonian}. Without loss of generality, we consider transport of the charge given by diag$(1,0,-1)$, i.e. the spin-1 $S^z$ operator. We consider the corresponding analogs to Eqs. \ref{initial state} and \ref{mag transfer} for the initial state and measured magnetization transfer. As expected, we find that the $t_*\sim U^{-6}$ behavior holds for this model as well, suggesting that it is in fact universal even in the presence of multiple quasiparticle types (Fig. \ref{fig:su3_U}).

\section{Discussion}
Using state-of-the-art MPS simulations, we have characterized the timescale for the breakdown of superdiffusive transport in integrable quantum spin chains. We have focused on a ladder configuration as a minimal non-integrable correction to the idealized isotropic Heisenberg chain, considering various perturbations that can be added in this geometry. In particular, we have confirmed the important distinction between integrability-breaking perturbations that preserve or break the underlying non-abelian symmetry from which superdiffusion arises. The timescale of diffusion onset is found to go as $t_*\sim \lambda^{-\alpha}$, with $\alpha=2$ (6) for symmetry-breaking (-preserving) perturbations, in agreement with recent results \cite{mccarthy_slow_2024}. Consistent results are found when considering another integrable point of the $S=1/2$ ladder with emergent $SU(4)$ symmetry, as well as in a generalization of the model to an $S=1$ ladder with $SU(3)\times SU(3)$ symmetry.

These different scaling laws could be helpful in interpreting observed corrections to superdiffusion in experiments, particularly those on solid-state systems in which integrability-breaking corrections are non-negligible. Since such experiments are not carried out at infinite temperature, they might also allow studies of how the onset of diffusion is modified by temperature. Ref. \onlinecite{mcroberts_parametrically_2024} considered the temperature dependence of the classical Heisenberg chain, mapping to the same ansatz for the crossover form of $D(t)$ due to two-soliton collisions. That ansatz leads to the scaling $t_*\sim \beta^8$ for the crossover from superdiffusion to diffusion, coming after the ballistic to superdiffusion crossover for which $t_*\sim \beta$ (for $\beta>1/J$), as in the quantum case~\cite{dupont_spatiotemporal_2021}; such a long time scale at low temperature would be difficult to access with fully quantum simulations.

We conclude that superdiffusion in integrable quantum spin chains has a surprising longevity even at high temperature, particularly to perturbations that respect the non-abelian symmetry, and that this longevity may be part of the reason why experiments searching for superdiffusion were rapidly successful. One can hope that other novel regimes of quantum transport will be similarly robust and observable.

\begin{acknowledgments}
We are grateful to Romain Vasseur for helpful discussions. This work is supported by the Quantum Science Center (QSC), a National Quantum Information Science Research Center of the U.S. Department of Energy (DOE). This research used resources of the National Energy Research
Scientific Computing Center, a DOE Office of Science User Facility
supported by the Office of Science of the U.S. Department of Energy
under Contract No. DE-AC02-05CH11231 using NERSC award
BES-ERCAP0032440. The code for simulations was built using the ITensors library \cite{fishman_itensor_2022}.
\end{acknowledgments}

\bibliography{addl_references,references}
\end{document}